\documentclass [12pt,preprint]{aastex}
\begin{document}

\title{Early Cosmic Chemical Evolution: Relating the Origin of a
Diffuse Intergalactic Medium and the First Long-Lived Stars}

\author{F.D.A. Hartwick}

\affil{Department of Physics and Astronomy, \linebreak University of 
Victoria,
Victoria, BC, Canada, V8W 3P6}
\begin {abstract}

Nucleosynthetic signatures in common between the gas responsible for the
high redshift Lyman $\alpha$ forest and a subsample of extremely metal
poor stars are found.  A simple mass loss model of chemical evolution with
physically motivated parameters provides a consistent picture in which the
gas is identified with that lost by supernova-driven winds during the first 
generation of
star formation.  Substantial mass loss occurs which can account for a
diffuse IGM with up to $80\%$ of the total baryon content and a peak
[C-O/H] abundance of $\sim-2.9$.  This mass loss component differs from one
produced later during galaxy formation and evolution that contributes to
a circum-galactic medium (CGM).  The CGM was shown earlier to 
have a mass of $\sim10\%$ of all baryons and peak [Fe/H]$\sim-1$. 

\end{abstract}

\keywords{intergalactic medium}

\section {Introduction}

It is well known that the stellar content of the universe is only about
$\sim10\%$ of the total baryonic content (e.g. Fukugita \& Peebles 2004).
It is also well established that vigorous star formation, which occurs
during galaxy formation and evolution, is accompanied by mass loss (e.g.
Veilleux et al. 2005).  According to some models this mass loss can account
for another $\sim10\%$ of all baryons (e.g. Hartwick 2004, hereafter H04).
What is the origin and composition of the remaining $\sim80\%$ of all
baryons? This question is addressed here using recent observational data
and simple chemical evolution arguments. 

Generally the abundance distribution exhibited by stellar spectra reflects
the material that formed the stars.  If one were to observe spectra of the
oldest stars (most metal poor?) in our Galaxy, then one expects to have
to look at high redshifts in order to find gas with the same
nucleosynthetic signature. 

At redshifts of order 3, the baryon budget is believed to be dominated
by the tenuous gas responsible for producing the Lyman $\alpha$ forest
observed in absorption in high redshift quasar spectra (e.g. Rauch 1998).
As discussed below, recent high quality data from the largest telescopes
combined with sophisticated reduction techniques have allowed the
detection of the heavy elements C, O, and Si in low column densities of 
hydrogen.  These new observational results have prompted a re-examination of a
possible connection between this gas and the earliest star formation
(Pop~III?) phase.  By modelling the observed distribution of carbon or
oxygen in this gas, we can predict the distribution of these elements and
characteristic abundance ratios expected in long-lived stars, if they had
been formed at the same early time.  Unfortunately both the gas and
stellar samples, are small and incomplete.  Yet, we find encouraging
consistency with the hypothesis when comparing these observations
with the results from a simple, physically motivated, mass loss model of
chemical evolution.  An important and surprising conclusion
is that long-lived stars with nucleosynthetic signatures in common 
with the Lyman $\alpha$ forest gas do exist. 

\section{The Observations}

\subsection{The Diffuse Gas}

Early work on the detection of carbon in low column density Lyman $\alpha$
clouds was carried out by Cowie, Songaila, Kim \& Hu (1995).  More recently, 
Simcoe, Sargent \& Rauch (2004) used survival analysis (a technique which makes
use of non-detections as well as detections) to construct the distribution
of [O/H] and [C/H] in low column density absorption lines along lines of
sight to seven quasars.  The data become incomplete at [O/H]$\lesssim-3.0$
(no floor was detected), but above that the distribution was log-normal
with mean [C-O/H]$\sim-2.85$ and dispersion $\pm0.75$, with these results
depending on the assumed shape of the UV background (UVB).  A second well
fit model used a softer UVB resulting in medians of [C/H]$\sim-3.1$ and
[O/H]$\sim-2.7$.  The ionizing UVB models are those of Haardt \& Madau
(1996) and are made up of contributions from a quasar continuum and a galaxy 
contribution.  A softer spectrum implies more contribution
from galaxies and less from quasars. 

The information in Fig.\ 14 of Simcoe, Sargent \& Rauch (2004) is the
starting point for the present work.  A second important observational
constraint comes from Schaye et al.\ (2003) who determined the
silicon to carbon ratio [Si/C] in the Lyman $\alpha$ forest down to very
low column densities.  They found this ratio to be $\sim0.77$, again
depending on the shape of the UVB that is assumed, with a softer UVB (more
galaxy and less quasar contribution) resulting in a lower [Si/C] ratio. 
The authors concluded that this [Si/C] ratio is unlikely to be lower than
$\sim0.5$. 

\subsection{The Stellar Sample}

When the chemical evolution model is constrained by observations of the
diffuse gas, one can predict the [C-O/H] distribution in stars, assuming
that luminous long-lived remnants were produced.  Further, the stars should
also show appropriate [Si/C] signatures.  If such stars were formed, we can
reasonably expect to find them among the various samples of extremely
metal poor stars which have been spectroscopically analyzed. The data 
discussed below follows from 
earlier work by Gratton \& Sneden (1988, 1991), McWilliam et al.\ (1995), 
Ryan, Norris \& Beers (1996), and others. 

Here we are concerned primarily with the abundances of C, O, Mg, Si, and
Fe.  Unfortunately, not all of these abundances are available from every
study, and care must be exercised when interpreting and combining these
results, since the samples are subject to strong observational selection
effects.  In addition, most of the work has been done on giant stars whose
atmospheres are subject to non-LTE effects, possible mixing effects, 1-D
versus 3-D effects, and incompleteness due to the weakness of lines
analyzed.  Carbon and oxygen are particularly vulnerable.  Our main
sources of data are the work of Cayrel et al.\ (2004) and a further
discussion of the carbon and oxygen abundances of the same stars by Spite
et al.\ (2005).  For carbon [C/Fe], the sources are the unmixed stars of
Spite et al.\ (2005), Akerman et al.\ (2004), as well as those stars from
Honda et al.\ (2004) and Barklem et al.\ (2005) with log
L/L$_{\odot}\lesssim2.3$ (calculated from log T$_{eff}$, log g, and
assumed mass of 0.8~M$_{\odot}$).  This selection criterion was used in order
to reduce the possibilty of including mixed-CNO atmospheres.  Note that all
sources, except Barklem et al., avoided including carbon-rich stars.  Data
for oxygen [O/Fe] are from Akerman et al.\ (2004) and Spite et al.\
(2005).  Following Spite et al., their [O/Fe] data were lowered by 0.25 dex to
agree
with the Akerman et al.\ data.  Magnesium abundances [Mg/Fe] are available
from all sources except Akerman et al., silicon [Si/Fe] from all sources
except Akerman et al.\ and Barklem et al., while values of [Fe/H] are
available for all stars. 

Generally, the above samples contain some of the most metal poor stars known. 
Most of these stars are more metal poor than the metal poorest globular 
clusters ([Fe/H]$\sim-2.4$, [O/H]$\sim-1.8$). As such they are assumed to have 
comparable or older ages ($\sim13-14 \times10^{9}$ yrs) and hence masses 
$\sim0.8$M$_{\odot}$.

The data are presented as plots of [O/H] and [C/H] versus [Fe/H], [Si/H],
and [Mg/H] in Figs.\ 1--3.  The solid line in each figure shows the
Population II sequence.  It is defined by the apparent clumping of many of the
stars (especially at the metal rich end) and from the figures, by the following
implied abundance ratios (i.e. [C/Fe]$\sim0.25$,
[O/Fe]$\sim0.65$, [Mg/Fe]$\sim0.30$, and [Si/Fe]$\sim0.35$). These ratios are 
similar to those found in earlier studies of the less metal poor Pop~II stars 
(e.g. the works cited in the first paragraph of this section).
The dashed lines delineate second
sequences where stars with nucleosynthetic signatures in common with the
Lyman $\alpha$ forest gas lie. In
every figure, the Cayrel et al.\ stars labelled 5, 30, 32, and 34 are
offset from the Pop~II sequence and straddle the dashed line.  Other stars
may also belong to this group, but either one or both of carbon, oxygen or
silicon abundances are not available to confirm this possibility or they are 
masked by observational uncertainty especially near the observational limit. 
The [C-O/H] abundances of these 4 stars are plotted as a histogram in the
right hand panel of Fig.\ 4.  Note that compared to Pop II stars, these stars 
have reduced carbon and oxygen but enhanced silicon with respect to iron. In 
particular, we note that the C and O independent ratio [Si/Fe] 
differs by 0.2 dex between the two groups.  
From Fig.2 the enhanced [Si/C] ratio is
$\sim0.6$.  Note also that the [O/C] ratio for these stars is $\sim0.4$,
similar to that of the Pop~II star sequence.  Based on the similarity of
the observed [C-O/H] distribution and [Si/C] ratio with that of the
diffuse gas, it is suggested that these 4 stars, and possibly several
others, are first generation stars. Note that because the model distributions 
in Figure 4 begin with an assumed oxygen abundance of zero and end when the 
Pop~II stars start to dominate, the distributions and the appropriate 
abundance ratios define what is referred to as first generation.

From Figs. 1--3 the first stellar
generation nuclear yields [C/Mg], [O/Mg], [Si/Mg], and [Fe/Mg] are $-0.4$, 
0.0, 0.2, and $-0.35$ respectively, with an estimated uncertainty in each of 
$\pm0.1$. 
These yields can be compared to those calculated by Heger \& Woosley (2002) 
for a variety of Population III enrichment scenarios that are shown 
graphically in 
their Figs. 3--5. The closest fit to the first three observed ratios above 
is the model represented by the solid line in their Fig. 5 (i.e. high energy 
explosions of 12--40 M$_{\odot}$ 
stars including exploding very massive stars ($\sim140-260$ M$_{\odot}$) 
modelled with a Salpeter-like IMF. The predicted ratios [C,O,Si/Mg] are 
$-0.31$, 0.03, and 0.48. Both carbon and especially silicon are over produced, 
but a very massive star contribution appears to be required.

\section { The Chemical Evolution Model}

The model employed here is similar to the one used to describe the global
star formation history (H04), and it is based on the simple mass loss
model (e.g. Hartwick 1976, also see Pagel 1997, and Binney \& Merrifield
1998).  Its novelty is a premature halting of the chemical evolution in a
distributed but physically motivated way.  Consider the standard one zone
mass loss model of chemical evolution.  Gas is slowly turned into stars
while being gradually enriched in heavy elements according to a specified
chemical yield, p.  Similtaneously, as a result of supernova driven winds,
gas with the same abundance as the gas currently forming stars is also being 
lost to further star formation at a rate which is proportional to the star 
formation rate. The `effective' yield (p$_{eff}$) determines the constant of 
proportionality, c=(p/p$_{eff}$)$-1$. 

If chemical evolution is suddenly halted before all of the gas is
exhausted, a discontinuity occurs in the metallicity distribution of stars 
formed as well as in the lost gas, leaving a reservoir of gas with
uniform composition available for later star formation.  The stellar
metallicity distribution can be described mathematically as the
undisturbed distribution multiplied by a complementary Heaviside unit step
function.
 
If the step function (or gas starvation function, f) is replaced with one
distributed in metallicity (i.e. we round the corners of the step function), 
then the one zone model appears as though it
contains many individually evolving boxes.  In H04, for example, the
complementary error function replaces the Heaviside function to be
consistent with the Gaussian distribution of stoppages assumed to occur as
a result of collisions among the numerous individual clumps. To provide more 
context for the model we give a brief summary of its application in H04.

There, by confining attention to a representative volume of the universe, 
the chemical evolution and star formation history associated with galaxy 
formation and evolution is derived from observations of stars and clusters 
in the Galaxy and in M31. The picture considers the 
anisotropic collapse of a number of star forming clumps. The gas in these 
clumps is assumed to have been enriched by a previous generation of star 
formation. 
The first collapse, perpendicular to the eventual rotation axis, results in 
collisions between low angular momentum clumps (M$_{t,blue}$) which terminates 
the star formation and is assumed to create the metal poor (blue) globular 
clusters. The stars already formed (M$_{s,blue}$) then constitute the extended 
metal poor 
halo and the gas released in the collisions (M$_{ml,blue}$) falls to the center
to form the bulge. The higher angular momentum clumps (M$_{t,red}$) continue 
to form stars and become more metal rich until they too begin to collide as 
they fall along the rotation axis. These collisions again terminate star 
formation and give rise to the metal rich (red) globular clusters. The stars 
already formed (M$_{s,red}$) are released to form a metal rich spheroid 
population, and the 
gas (M$_{ml,red}$) dissipates to form the disk.
Interestingly, in this picture only $\sim20$\% of the available baryons are 
required to provide an acceptable fit to the observed cosmic star formation 
rate density. Further, before 
collisions terminate star formation within the clumps, 
approximately half of these baryons are returned to the IGM by supernova 
driven winds (designated (M$_{WHIM,blue}$+M$_{WHIM,red}$) in H04 but 
renamed here M$_{CGM}$, the 
circum-galactic medium (CGM) component). Accounting for the `leftover' 80\% of
the baryons provided one of the main motivations for the present work.

Here we assume there are a large number of star forming clumps approaching
the transition to Pop~II as the carbon and oxygen abundances approach
[C-O/H]$\sim-3$.  At this point the gas can cool more efficiently which in
turn allows low-mass Pop~II stars to form more readily (e.g. Bromm \& Loeb
2003).  To characterize this transition we define f to be a complementary
extreme value or Fisher Tippett distribution (see equation (6) below). 
Because the present context of the model is different from that of H04, we
redefine the variables and rewrite the equations governing the chemical
evolution.  The conserved quantity is the total baryon mass, M$_{t}$.  It
is made up of 4 components: the gas contributing to star formation M$_{g}$
which initially is the same as M$_{t}$, the mass in long-lived stars
and/or remnants M$_{s}$, the gas lost due to supernova driven winds
identified with the diffuse intergalactic medium M$_{dIGM}$, and the gas
left to form later generations M$_{LG}$.  From H04, with the oxygen
abundance O replacing Z, M$_{dIGM}$ replacing M$_{WHIM}$, and M$_{LG}$   
replacing M$_{ml}$ the equations become: 

\begin{equation}
M_{t}=M_{g}+M_{s}+M_{dIGM}+M_{LG}
\end{equation}

\begin{equation}
M_{g}={M_{t}}\times{exp(-(\rm{O-O}_{0})/p_{eff})}\times{f}
\end{equation}

\begin{equation}
\frac{dM_{s}}{dlog \rm O}={ln10}\times {\rm O}\times {M_{g}}/p
\end{equation}

\begin{equation}
M_{dIGM}={({{p}/{p_{eff}}-1})}\times{M_{s}}
\end{equation}

 and

\begin{equation}
\frac{dM_{LG}}{dlog \rm O}={-M_{t}}\times{exp(-(\rm{O-O}_{0})/p_{eff})}\times{
\frac{df}{dlog \rm O}}
\end{equation}

where the complementary extreme value function is written

\begin{equation}
f=1-exp(-exp(-log(\rm{O/O_{f}}))/W)
\end{equation}

For this particular problem the initial conditions are M$_{g}$=M$_{t}$,
M$_{s}$=0, M$_{dIGM}$=0, M$_{LG}$=0, and O$_{\rm 0}$=0. The parameters O$_{f}$ 
and W in equation (6) represent the point of inflection and the `width' of the 
transition from one to zero of the extreme value distribution \linebreak (see 
e.g. http://mathworld.wolfram.com/ExtremeValueDistribution.html).

\subsection{Model Results}

As shown in Table 1, 4 input parameters are required to construct a
model.  The constraints considered are a given true yield p for oxygen, a
fit of the distribution of M$_{dIGM}$ to the high [O/H] end of the Simcoe
et al.\ histogram (Fig.\ 4), and that M$_{LG}$ be 19\% of the total baryon
content as required by the successful model of cosmic chemical evolution
in H04.  The oxygen yield was determined by combining the [O/Fe] value of
0.35 from Fig.\ 1 with the yield assumed in H04 of [p$_{\rm Z}$/H]$=-0.11$ 
(representing the iron yield) 
to give [p$_{\rm O}$/H]$=0.24$. Note that while Z was used as the model 
variable in H04 its observational counterpart there was always assumed to be 
[Fe/H]. 
The remaining 3 input parameters were then
varied in order to satisfy the last two constraints resulting in the
`nominal' values in Table 1.  In order to gauge the robustness of the
results, each input parameter was varied separately by the amount
tabulated, and its effects on each of the output values are also shown.
The top row for each output variable shows the result of a positive
variation, and the bottom row shows the negative variation.  For example, 
the nominal value of M$_{dIGM}$/M$_{t}$ is 0.81. If [p$_{\rm O}$/H] is 
increased by 0.3 M$_{dIGM}$/M$_{t}$ remains unchanged at 0.81. If 
[p$_{\rm O}$/H] is 
decreased by 0.3, M$_{dIGM}$/M$_{t}$ also remains unchanged at 0.81. However 
[O$_{eff}$/H] is increased by 0.3, M$_{dIGM}$/M$_{t}$ becomes 0.69 and it is 
increased to 0.9 for a decrease in [O$_{eff}$/H] of 0.3 etc. In order to
compare with the observations, the calculated distribution in the left
hand plot of Fig.\ 4 was smoothed with a Gaussian kernel of width $\pm0.5$
and the stellar distribution with a $\pm0.2$ kernel to allow for
observational uncertainties.  The two calculated distributions appear to be
quite consistent with the (limited) observations. The [O/H] distribution of 
gas from which later stellar generations (Pop~II and beyond) will form is 
given by equation (5).

One interesting output parameter is the value of M$_{s}$/M$_{t}$ (i.e. the
first generation fraction of long-lived stars and remnants, fg).  Assuming 
that this
number is representative of the universe as a whole, then
$\Omega_{s,fg}/\Omega_{b}=8.3 \times10^{-4}$.  If we further assume that a
Magorrian-like relation holds between black hole mass and stellar mass
(i.e. M$_{\bullet}$/M$_{\star}\sim0.002$), then from the WMAP value of the 
baryon density ($\rho_{b}h^{2}=0.0224$, Spergel et al.\ 2003), one obtains 
$\rho_{\bullet,fg}\sim1.0\times10^{4}$ M$_{\odot}$/Mpc$^{3}$.  It is
interesting to compare this number to the Yu \& Tremaine (2002)
determination of the same statistic at redshift zero:
$2.5\pm0.4\times10^{5}h^{2}_{0.65}$ M$_{\odot}$/Mpc$^{3}$.  If the black
hole seeds were formed during this first generation of star formation,
then they had to have grown by $\sim25$ times between then and the
present time under our assumed cosmogony and implied `normal' IMF. However,
the nuclear yields discussed earlier suggest that the first generation IMF 
extends to very massive stars. This should result in additional massive 
black holes being formed and make the above growth factor an upper limit.

The value of the output parameter, M$_{LG}$/M$_{t}$, was constrained to be 
the baryon fraction that was determined in H04 to successfully describe the
cosmic star formation history through galaxy formation and evolution. (From 
Table 2 of H04 $\Omega_{b}=\Omega_{\star}+\Omega_{WHIM}+\Omega_{diffuse}$. 
Hence the constrained value of 
M$_{LG}$/M$_{t}$ is ($\Omega_{\star}+\Omega_{WHIM})/\Omega_{b}=0.19$ with an 
estimated uncertainty of $\pm0.04$) This 
number includes the mass that is lost during this process, and is the 
component that we 
now refer to as the circum-galactic medium (CGM but called WHIM in H04)
in order to distinquish it from the dIGM discussed in this paper. In H04 it 
was found that 
M$_{CGM}$/M$_{t}\sim0.09$ with a heavy element content
[Fe/H]$\sim-1$.  Thus, we expect there to be two gaseous components. As shown 
in Table 1, the dIGM is much more massive (M$_{dIGM}$/M$_{t}\sim0.81$) and 
much more metal poor ([O/H]$\sim-2.9$).

If long-lived stars are formed during the Pop~III phase, one may ask 
where are the stars with [Fe/H]$<-6$. The answer is that such stars would be 
exceedingly rare. In a 
sample of 10,000 first generation stars only 15 are predicted to have 
[Fe/H]$<-6$ ([O/H]$<-5.65$) based on the model distribution in Fig.4. 
Currently 2 stars are known with [Fe/H]$<-5$: HE 0107-5240 and HE 1327-2326 
(e.g. Aoki et al.\ 2005). However, both of these stars show very large over 
abundances of carbon and so are apparently different from the stars discussed 
above.

\section{Discussion}

By identifying nucleosynthetic signatures in common, a case has been made 
for a connection between the diffuse IGM, as identified at high
redshift with the Lyman $\alpha$ forest, and the first stellar generation as 
represented by a subsample of extremely metal poor stars.
This is clearly a first step as both sets of data are small and
incomplete. Some theoretical arguments suggest that only high mass
stars with no long-lived luminous stellar remnants are made during the 
Population~III phase (e.g. Bromm \& Larson 2004) while others (e.g. Nakamura 
\& Umemura 2001) have  
argued that the first generation IMF is bimodal with peaks at $\sim100$ and 
$\sim1$ M$_{\odot}$. Our result suggests that 
nucleosynthesis from intermediate mass stars as well as very massive stars is 
required to satisfy the yields found in \S2.2. These nuclear yields, as well as
others 
available but not discussed here, should allow much tighter constraints to be
placed on the Pop~III IMF and the accompanying ratio M$_{\bullet}$/M$_{\star}$.

Other extensions to this work are obvious.  Every effort should be made to
measure both oxygen and silicon in the burgeoning samples of extremely
metal poor stars.  Work should be concentrated on the highest gravity stars
to circumvent problems with CNO mixing which occurs in the most luminous
stars.  In the case of the Lyman $\alpha$ forest observations, it would be
useful to investigate the ionizing UV background to see if it could be
adjusted consistently to similtaneously allow [O/C]$\sim0.4$ and 
[Si/C]$\sim0.6$ in the
gas as is observed in the stellar sample. 

Finally, our simple cosmogony requires two gaseous components.  A
circum-galactic one (CGM) arising from mass loss associated with galaxy
formation and evolution, whose mass is $\sim10\%$ of all baryons, and a
diffuse IGM which contains $\sim80\%$ of the total baryon content but
which is also much more metal poor than the CGM.  However, the model makes
no predictions about the thermodynamic state of the gas or how it evolves.
Rather, sophisticated cosmological simulations follow the evolution of
these baryons which are presently considered to be in the form of a
filamentary web of very low density gas (e.g. Cen \& Ostriker 1999).  One
aspect of the model, which may be relevant to the search for local
`missing' baryons (e.g. Nicastro et al.\ 2005) and that is independent of
the thermodynamic state, is that the $composition$ of most of the gas may
not have evolved and may still be peaked at [C-O/H]$\sim-2.9$.  This could
make it very difficult to observe with present instrumentation. 

\acknowledgments

The author wishes to acknowledge financial support from an NSERC (Canada)
discovery grant. 


\clearpage

\begin{figure}
\plotone{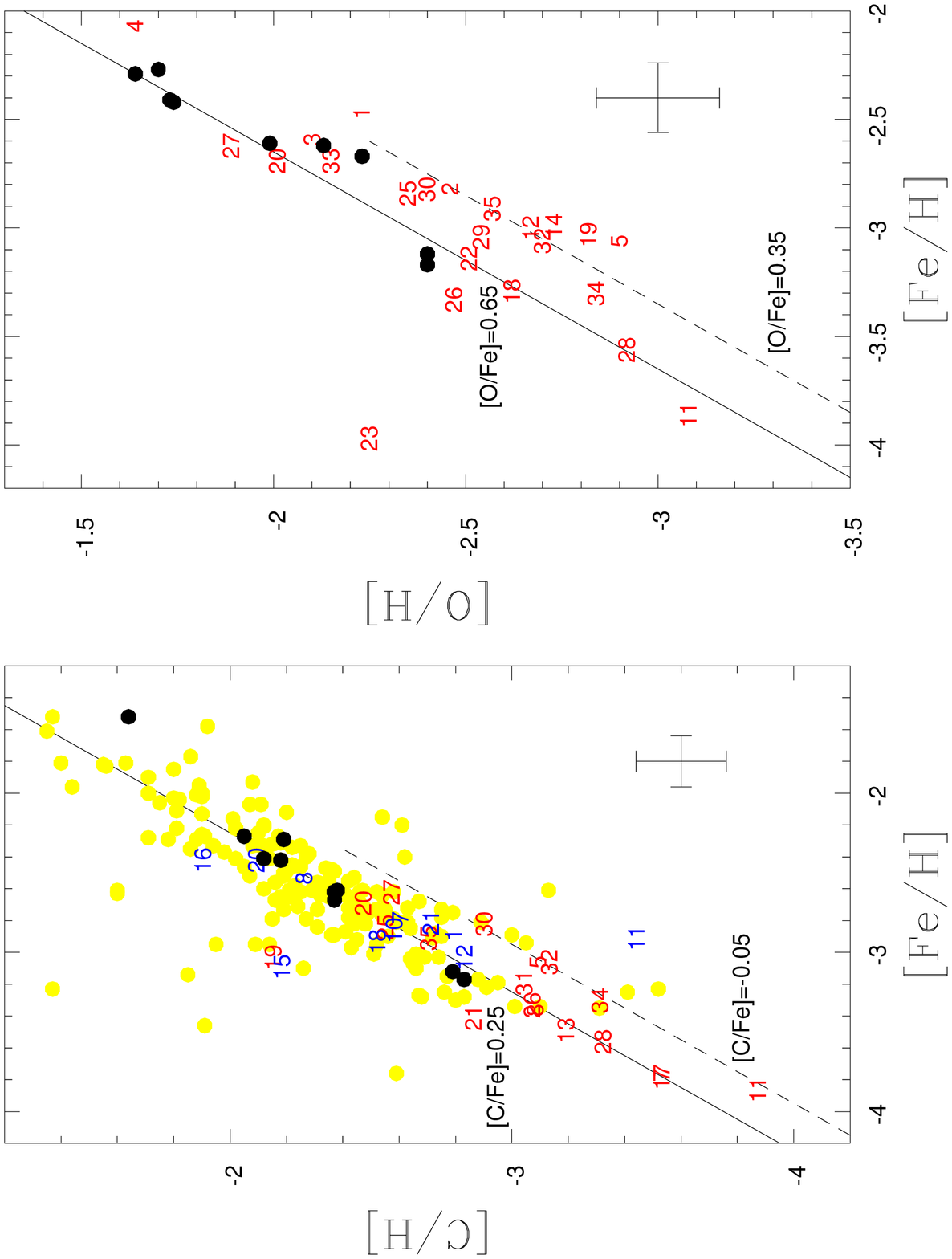}
\caption{ (left) [C/H] vs [Fe/H], (right) [O/H] vs [Fe/H] for extremely
metal poor stars.  Red numbers - data from Spite et al.\ (2005).  Blue
numbers correspond to stars in order of their appearance in the data
tables of Honda et al.\ (2004), black dots are data from Akerman et al.
(2004), and yellow dots are data from Barklem et al.\ (2005).  The solid line 
is the Population II sequence and the dashed line is the locus of stars with
abundances considered to be in common with the Lyman $\alpha$ forest. Typical 
error bars are shown.} 
\end{figure}

\clearpage

\begin{figure}
\plotone{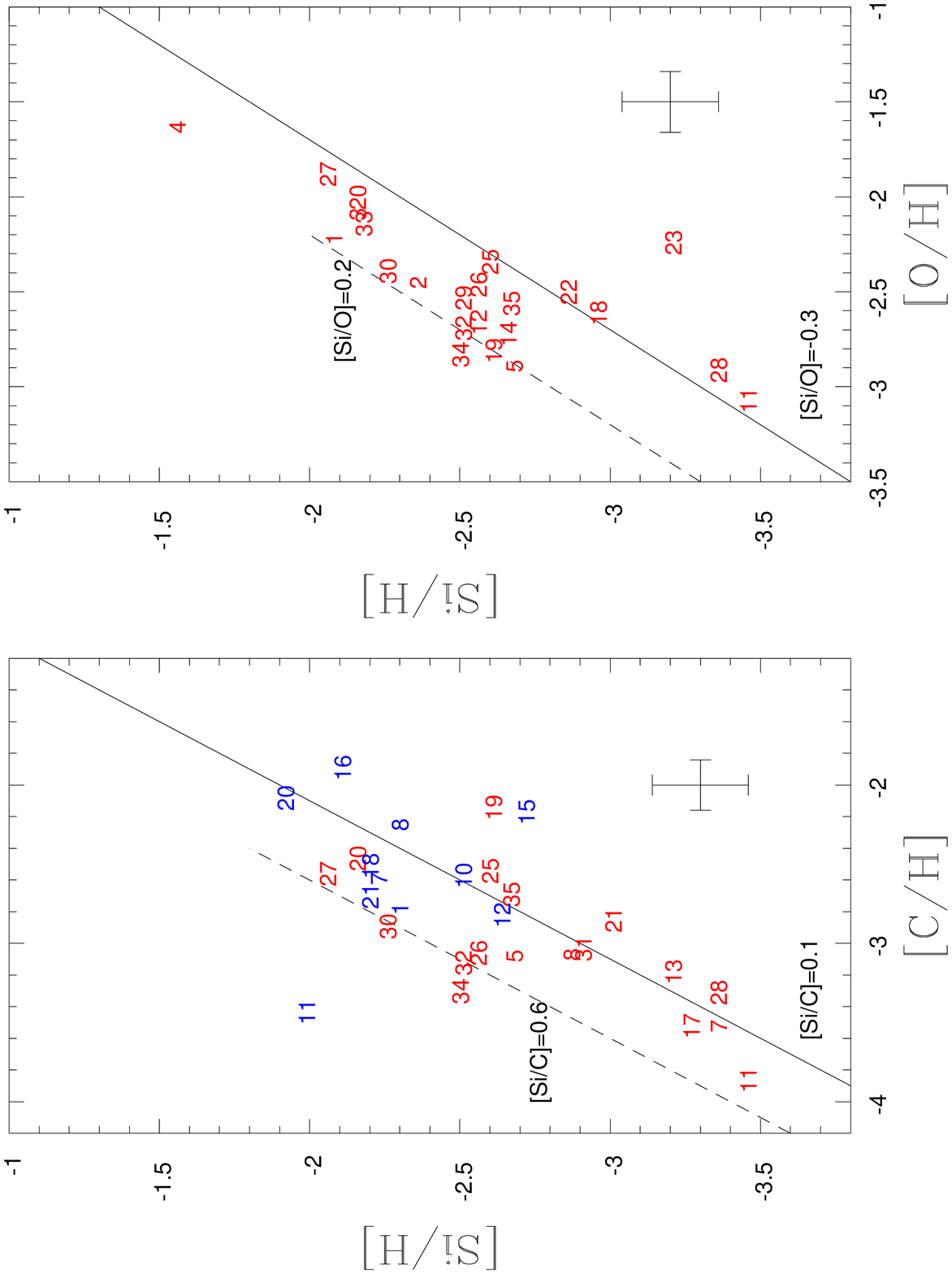}
\caption{The same as Fig.\ 1 except for [Si/H] vs [C/H] and [Si/H] vs [O/H]}
\end{figure}

\clearpage

\begin{figure}
\plotone{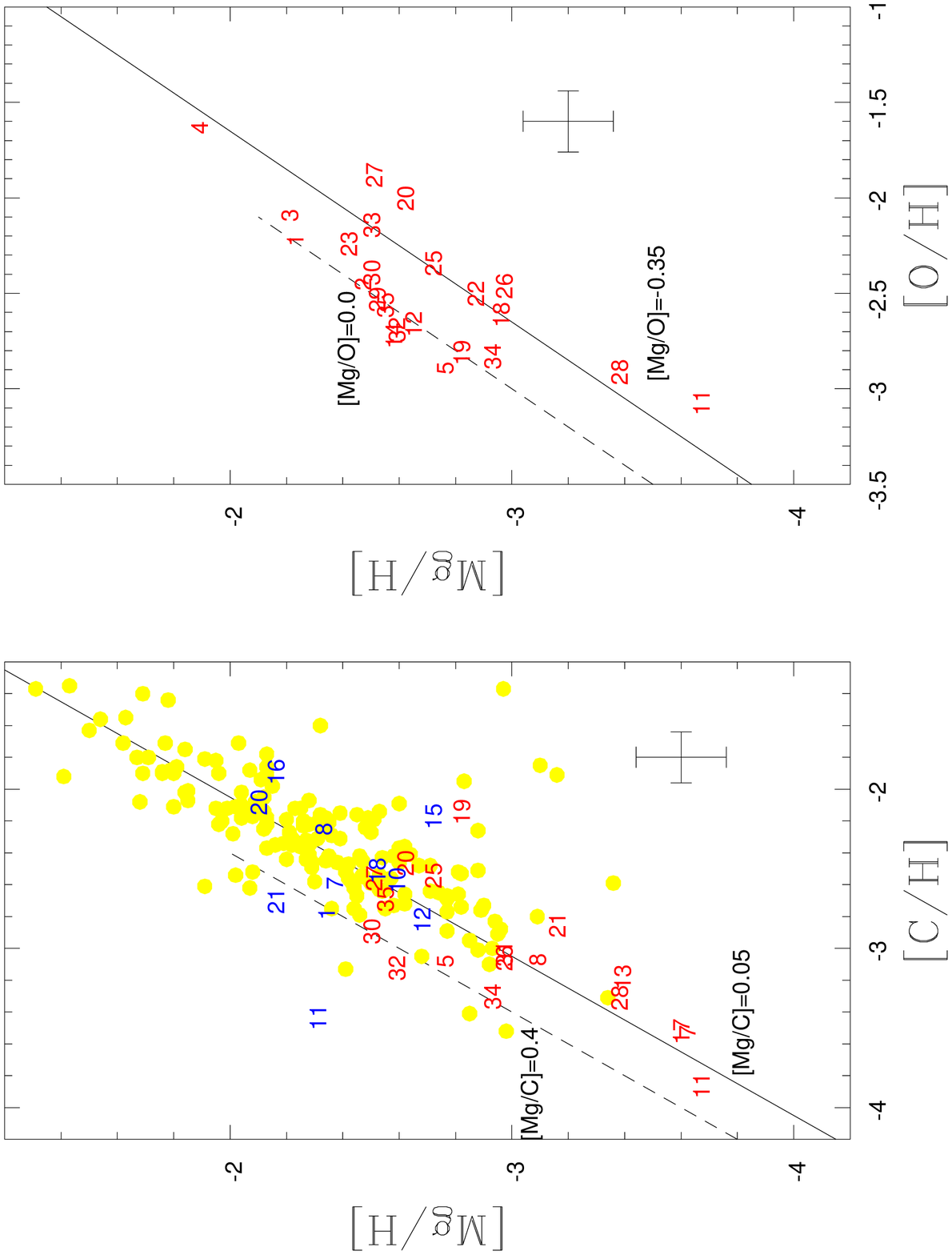}
\caption{The same as Fig.\ 1 except for [Mg/H] vs [C/H] and [Mg/H] 
vs [O/H]}
\end{figure}

\clearpage

\begin{figure}
\plotone{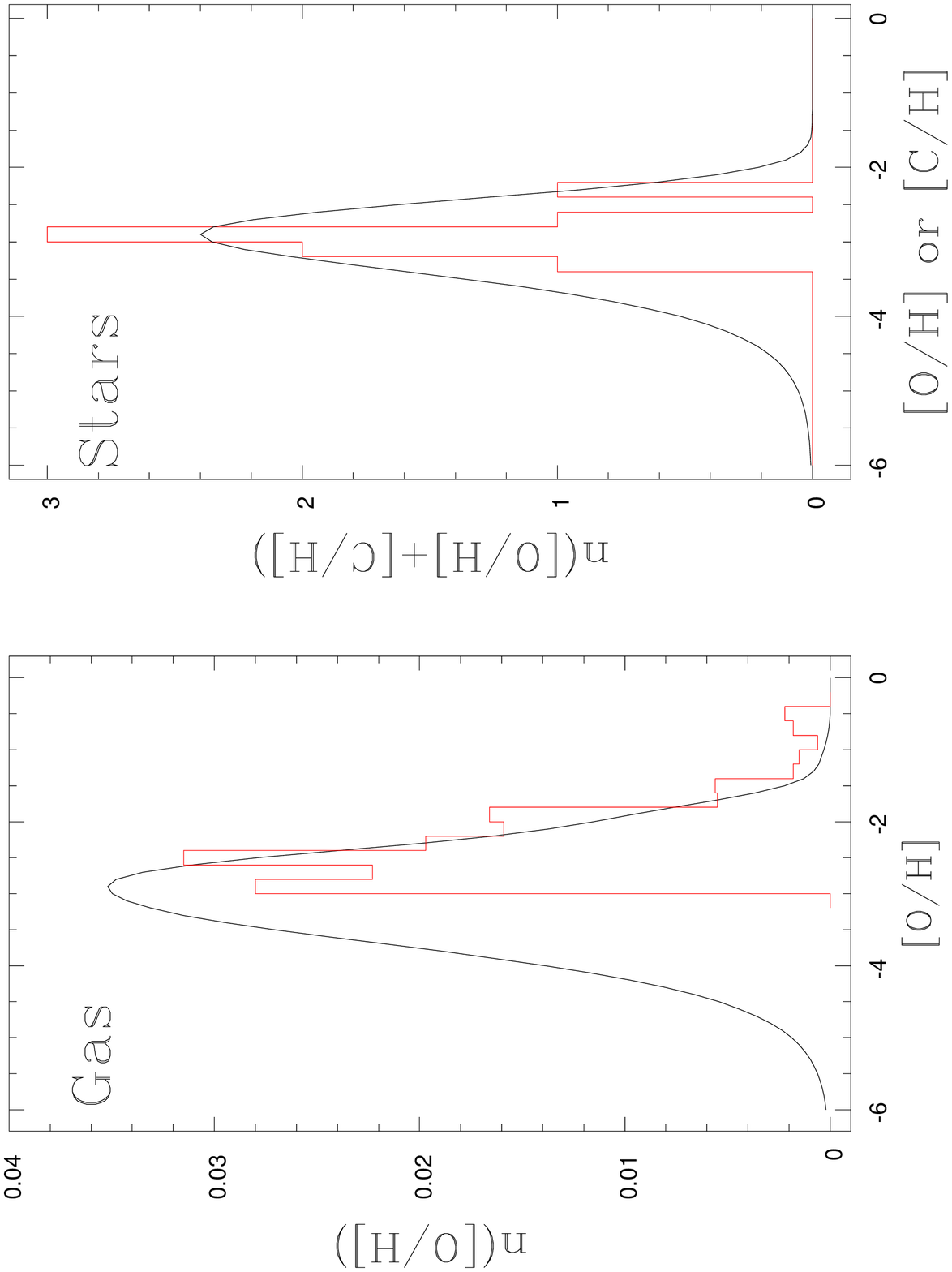}
\caption{(left) A comparison of the observed abundance distribution of
[O/H] in the low column density Lyman $\alpha$ forest from Simcoe et al.
(2004), with the mass loss component from the model (solid line).  (right) 
The solid line 
shows the predicted distribution of [O/H] for the stellar component of the 
model. The histogram shows the distribution of [C/H] $and$ 
[O/H] for the 4 stars considered prime first
generation candidates.  The normalization of the ordinate in both figures
is arbitrary.} 
\end{figure}

\clearpage

\begin{deluxetable}{cccccc}
\tablenum{1}
\tablecolumns{6}
\tablecaption{Model Parameters and Changes due to Variations in Input
Parameters\tablenotemark{a}}
\tablehead{Parameter & Input &
\colhead{[p$_{\rm O}$/H]} &
\colhead{[O$_{eff}$/H]} &
\colhead{[O$_{f}$/H]} &
\colhead{1/W} \\
\colhead{Output} &
\colhead{Nominal value} &
\colhead{$0.24\pm0.3$} &
\colhead{$-2.75\pm0.3$} &
\colhead{$-2.50\pm0.3$} &
\colhead{$1.5\pm0.5$}
}
\startdata
& & $4.16\times10^{-4}$ & $1.42\times10^{-3}$ & $9.24\times10^{-4}$ &
$8.51\times10^{-4}$  \nl
   M$_{s}$/M$_{t}$ & $8.30\times10^{-4}$ & \nl
& & $1.66\times10^{-3}$ & $4.63\times10^{-4}$ & $7.10\times10^{-4}$ &
$7.95\times10^{-4}$  \nl
\nl
& & $0.81$ & $0.69$ & $0.90$ & $0.83$ \nl
   M$_{dIGM}$/M$_{t}$ & $0.81$ & \nl
& & $0.81$ & $0.90$ & $0.69$ & $0.78$ \nl
\nl
& & $0.19$ & $0.31$ & $0.10$ & $0.17$ \nl
   M$_{LG}$/M$_{t}$ & $0.19$ & \nl
& & $0.19$ & $0.10$ & $0.31$ & $0.22$ \nl
\nl
& & $-2.90$ & $-2.70$ & $-2.85$ & $-2.90$  \nl
   {[O/H]$_{max}$} & $-2.90$ & \nl
& & $-2.90$ & $-3.15$ & $-2.95$ & $-2.90$ \nl
\nl
\enddata
\tablenotetext{a}{For each output parameter, the upper row gives the
value due to the indicated positive variations in each input parameter
separately.  The lower row are results from a negative variation. See text.}
\end{deluxetable}

\end{document}